# Element-specific ultrafast lattice dynamics in FePt nanoparticles

Diego Turenne[1], Igor Vaskivskiy[1,2], Klaus Sokolowski-Tinten[3], Xijie Wang[4,5,6], Alexander H. Reid[7], Xiaoshe Shen[7], Ming-Fu Lin[7], Suji Park[7], Stephen Weathersby[4], Michael Kozina[7], Matthias Hoffmann[7], Jian Wang[8], Jakub Sebesta [1], Yukiko K. Takahashi[8], Oscar Grånäs[1], Peter Oppeneer[1], Hermann A. Dürr[1*]

[1] Department of Physics and Astronomy, Uppsala University, Box 516, 75120 Uppsala, Sweden

[2] Department of Complex Matter, Jozef Stefan Institute, Jamova 39, Ljubljana SI-1000, Slovenia

[3] Faculty of Physics and Centre for Nanointegration Duisburg-Essen, University of Duisburg-Essen, Lotharstrasse 1, 47048 Duisburg, Germany

[4] Accelerator Division, SLAC National Accelerator Laboratory, 2575 Sand Hill Road, Menlo Park, CA 94025, USA

[5] Faculty of Physics, University of Duisburg-Essen, 47048 Duisburg, Germany.

[6] Department of Physics, TU Dortmund University, 44227 Dortmund, Germany.

[7] Linac Coherent Light Source, SLAC National Accelerator Laboratory, 2575 Sand Hill Road, Menlo Park, CA 94025, USA

[8] Magnetic Materials Unit, National Institute for Materials Science, Tsukuba 305-0047, Japan

*Corresponding Author E-mail: hermann.durr@physics.uu.se

ABSTRACT: Light-matter interaction at the nanoscale in magnetic alloys and heterostructures is a topic of intense research in view of potential applications in high-density magnetic recording. While the element-specific dynamics of electron spins is directly accessible to resonant x-ray pulses with femtosecond time structure, the possible element-specific atomic motion remains largely unexplored. We use ultrafast electron diffraction to probe the temporal evolution of lattice Bragg peaks of FePt nanoparticles embedded in a carbon matrix following excitation by an optical femtosecond laser pulse. The diffraction interference between Fe and Pt sublattices enables us to demonstrate that the Fe mean-square vibration amplitudes are significantly larger that those of Pt as expected from their different atomic mass. Both are found to increase as energy is transferred from the laser-excited electrons to the lattice. Contrary to this intuitive behavior, we observe a laser-induced lattice expansion that is larger for Pt than for Fe atoms during the first picosecond after laser excitation. This effect points to the strain-wave driven lattice expansion with the longitudinal acoustic Pt motion dominating that of Fe.

I. INTRODUCTION

Future magnetic data storage media will require magnetic nanoparticles with stable ferromagnetic order at diameters of only 10 nm and smaller [1]. In this respect, granular thin films of the L1$_0$-ordered phase of FePt displaying perpendicular magnetic anisotropy are one of the most suitable storage media. The FePt nanoparticles composing such granular materials remain ferromagnetic as a result of the strong magnetocrystalline anisotropy needed to overcome the superparamagnetic limit [2-5]. However, a byproduct of strong magnetocrystalline anisotropy is the large magnetic field required to reverse the nanoparticle magnetization. Applications strive to reduce the magnetic switching field by locally heating the nanoparticles above their Curie temperature with a laser in order to thermally assist the switching, a technique known as heat-assisted magnetic recording [6].

The magnetization dynamics of FePt nanoparticles following optical femtosecond (fs) laser excitation has been the subject of various studies resulting in the observation of sub-ps demagnetization [7, 8] and even all-optical magnetic switching [9]. However, much less is known about the ultrafast lattice response which could only recently be addressed using ultrafast x-ray and electron scattering. Reid et al. [10] showed that the response of suspended 13 nm FePt nanoparticles is characterized by a lattice expansion along the Fe and Pt layers (a, b direction in Fig. 1) accompanied by a contraction of the lattice spacing in the perpendicular direction (c direction in Fig. 1). This reflects a magnetostrictive stress on the lattice due to the laser-induced quenching of the magnetic order [10]. Key of such studies is that the nanoparticle lattice is free to follow the intrinsic stress buildup within the particles. For instance, FePt nanoparticles with their lattice spacing along the Fe/Pt layers locked into that of a supporting substate still react to magnetostrictive stress via a lattice contraction along the perpendicular direction [11].

Here we address the question if the observed changes of the FePt nanoparticle lattice are uniform for the Fe and Pt sublattices. Utilizing the constructive and destructive interference of scattering from both atomic sublattices for Bragg peaks with even and odd sums mof Miller indices, respectively, we show that Pt expands faster than the Fe sublattice. We correlate this observation with element-specific measurements of the temporal variations observed in mean square vibrations and Brillouin zone boundary phonon occupations.

II. EXPERIMENT

Single crystalline $L1_0$ FePt was grown epitaxially onto a single-crystal MgO(001) substrate by co-sputter Fe, Pt and C [12]. This resulted in FePt nanoparticles of approximately cylindrical shape with heights of 6 nm and diameters in the 4-12 nm range with an average of 7.1 ± 1.8 nm as corroborated with transmission electron microscopy. The FePt nanoparticles form with a and b crystallographic directions along the MgO surface. The volume in-between nanoparticles is filled with glassy carbon at 30% volume fraction. Subsequently the MgO substrate was chemically removed and the FePt-C films were floated onto copper wire mesh grids with 100 μm wide openings.

The dynamic lattice response of FePt was measured by ultrafast electron diffraction in a transmission geometry (see Fig. 1A) with 3.6 MeV electrons from the SLAC ultrafast electron diffraction source [13]. The pump-probe experiments described below were carried out at room temperature with 1.5 eV / 50 fs laser excitation at a nominal pump fluence of 4 mJ/cm$^2$. We note that this required a variation of the actual incident pump fluence depending on the sample tilt angles.

To meet the Bragg condition for different lattice reflections the film was rotated around axes normal to the probe beam. Due to geometrical reasons rotation angles were limited to 45° from normal incidence. Measurements made at normal incidence, with the [001] c-axis parallel to the electron beam, showed changes of the diffraction pattern displayed in Fig. 1B as the difference of the pattern at 1ps pump-probe time delay with respect to that obtained before time zero, i.e. when the optical pump pulses arrive after the electron probe pulses. The Bragg peak positions were determined using a fit of two-dimensional Gaussian profiles to the experiments [10]. While at normal electron incidence we probe Bragg peaks along the a,b-crystal axes (see inset of Fig. 1C), also c-axis Bragg peaks are accessible when the lattice was tilted away from [001] normal incidence direction. Following ref. [10] we collected different Bragg reflection position and intensity data after a rotation of the film about the [100] a-axis to a point where the [111] reflections were easily visible. The observed time evolution data of Bragg peaks with different projections along the out-of-plane direction is used to reconstruct the [001] c-axis Bragg intensity following ref. [10]. Results are shown in Fig. 1D together with the determination of the average unit cell size variation in Fig. 1C.

The data displayed in Fig. 1 resemble those obtained for larger FePt nanostructures [10]. They show an initial lattice expansion along the a and b-axes as well as a concomitant c-axis reduction of the lattice spacing due to the reduction of the magneto-strictive stress following a laser-induced ultrafast quenching of the ferromagnetic order [10]. This leads to a long-lasting average increase of the FePt unit cell during approximately 1 ps after laser excitation followed by a volume-conserving breathing mode at a frequency given by the time it takes an acoustic lattice strain wave to move through the nanoparticle [10].

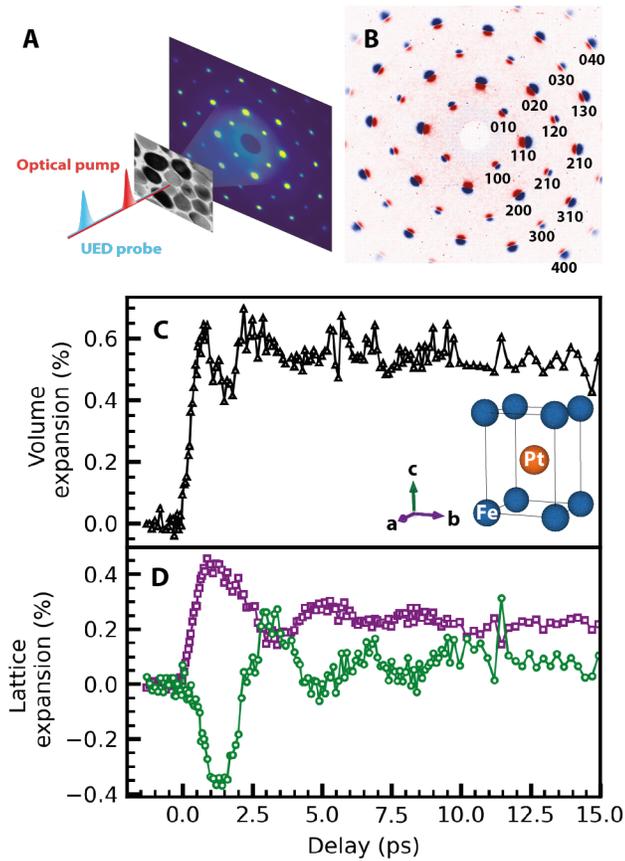

**Fig. 1. Average lattice expansion**. (**A**) schematic of the experimental UED pump-probe setup in transmission geometry with optical pump and UED probe beams incident along the FePt c-axis direction (see inset in panel C). (**B**) Difference between diffraction patterns at 1ps and at negative delay. The color coding is such that red/blue implies an intensity increase/decrease. (**C**) average lattice volume evolution (**D**) Temporal evolution of the in-plane (a and b crystallographic directions) lattice spacing in violet and the out-of-plane (c crystallographic direction) atomic spacing in dark green.

III. RESULTS:

A. Element-Specific mean square displacement dynamics

Mean square displacements are a way to measure vibrations of atoms around their equilibrium positions in a crystal, unveiling the energy stored in the lattice. Within the context of ultrafast lattice dynamics, they serve as an ultrafast proxy for the energy stored in the lattice, making them an ideal tool for tracking the energy flow in an excited, out-of-equilibrium system. The mean square displacements of atoms affect the scattering intensity through the Debye-Waller factor, $M$. In the case of FePt, the integrated intensity of the Bragg peaks is dependent on the Debye-Waller factors of both Iron, $M_{Fe}$, and Platinum, $M_{Pt}$. The Debye-Waller factor for diatomic species like FePt has been widely studied in a static regime [14-18], but their analysis of Debye-Waller factors in a pump-probe scheme remains constrained for either mono-atomic species [19, 20] or by using lattice temperatures without a clear distinction of the different chemical species [21-23].

Here, we separate the Debye-Waller factors for Fe and Pt atoms using the constructive and destructive interference of waves from both atoms depending on the selected Bragg diffraction order. For a Bragg peak with reciprocal lattice vector $\mathbf{q}_{hkl}$ the Bragg peak intensity $I_{hkl}$ is given by constructive and destructive interference between Fe and Pt depending on whether $h+k+l$ is even and odd, respectively, [24] i.e.

$$I_{hkl} = |e^{-M_{Pt}}F_{Pt}(\mathbf{q}_{hkl}) \pm e^{-M_{Fe}}F_{Fe}(\mathbf{q}_{hkl})|^2 \qquad (1)$$

where $M_{Fe,Pt} = -\frac{1}{2}\langle(\mathbf{q}_{hkl} \cdot \mathbf{u}_{Fe,Pt})^2\rangle$ and $\langle...\rangle$ denotes the Brillouin zone average. The large wavevector transfers in UED experiments enable observation of multiple Bragg peaks in the same experimental geometry. We can, therefore, use the measured Bragg intensities to separate the scattering contributions of Fe and Pt atoms in Eq. (1). The inset of Fig. 2A shows the $q$-dependence of $I_{hkl}$ measured in equilibrium, i.e. before any laser excitation occurs. If we interpolate $I_{even}$ and $I_{odd}$ to the same reciprocal lattice vectors, $q$, we can use Eq. (1) to obtain

$$e^{-M_{Fe}} 2I_e F_{Fe}(\mathbf{q}) = \sqrt{I_{even}} - \sqrt{I_{odd}} \text{ and}$$

$$e^{-M_{Pt}} 2I_e F_{Pt}(\mathbf{q}) = \sqrt{I_{even}} + \sqrt{I_{odd}} \quad (2)$$

Figure 2 shows the corresponding evaluation of the Debye-Waller factors $M_{Fe,Pt}$ from Bragg peak intensities in our UED data. We can eliminate the term $2I_e F_{Pt}(\mathbf{q})$ from Eq. (2) by normalizing to the values of $\sqrt{I_{even}} \pm \sqrt{I_{odd}}$ in equilibrium, i.e. before laser excitation. For each time delay (delays of 0.2 and 3 ps are shown in Figs. 2A, B) we obtain $\Delta M_{Fe,Pt} = -\frac{1}{2} q^2 \Delta \langle (\boldsymbol{\varepsilon}_{hkl} \cdot \mathbf{u}_{Fe,Pt})^2 \rangle$, i.e. the change of $M_{Fe,Pt}$ relative to equilibrium, as the slope of a log-plot of $\sqrt{I_{even}} \pm \sqrt{I_{odd}}$ normalized to their equilibrium values vs. $q^2$. Here $\boldsymbol{\varepsilon}_{hkl}$ is a unit vector pointing towards the Bragg peak of order $hkl$.

Figure 2C shows the determined changes of Fe (blue symbols) and Pt (orange symbols) mean square displacements, $\Delta \langle (\boldsymbol{\varepsilon}_{hkl} \cdot \mathbf{u})^2 \rangle$, projected onto the direction, $\boldsymbol{\varepsilon}_{hkl}$, of the respective reciprocal lattice vectors, $\mathbf{q}_{hkl} = q_{hkl} \boldsymbol{\varepsilon}_{hkl}$, versus pump-probe time delay. The observed increase of $\Delta \langle (\boldsymbol{\varepsilon}_{hkl} \cdot \mathbf{u})^2 \rangle$ can be described by two exponentials of the form $A_1(1 - e^{t/\tau_1}) + A_2(1 - e^{t/\tau_2})$. The determined fit parameters are summarized in Table 1.

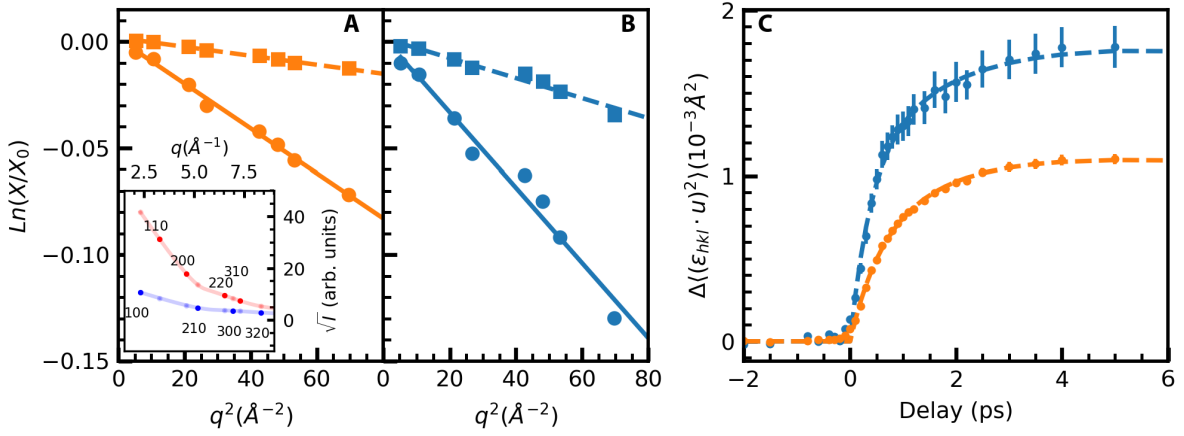

**Fig. 2. Debye-Waller dynamics**. (**A**) Semi-logarithmic plot of $X = \sqrt{I_{\text{even}}} + \sqrt{I_{\text{odd}}}$ versus $q^2$ as described in the text. $X_0$ denotes the values before laser excitation. The slope of the straight-line fits represents the change in Debye-Waller factor, $M_{\text{Pt}}$, at time delays of 0.2 ps (squares) and 3 ps (circles) relative to before laser excitation. The inset displays the measured intensities of the respective Bragg peaks for $h+k+l$ even (red line and symbols) and odd (purple line and symbols). The light symbols indicate the interpolated intensities as described in the text. (**B**) Same as in A) but for $\sqrt{I_{\text{even}}} - \sqrt{I_{\text{odd}}}$ resulting in the Debye-Waller factor, $M_{\text{Fe}}$. (**C**) Temporal evolution of the change in mean-square displacements, $\Delta\langle(\boldsymbol{\varepsilon}_{hkl} \cdot \mathbf{u})^2\rangle$, projected onto the direction $\boldsymbol{\varepsilon}_{hkl}$, of the reciprocal lattice vector $\mathbf{q}_{hkl}$. Shown are data for Pt (orange symbols) and Fe (blue symbols) with the fits described in the text indicated by dashed lines.

|  | $A_1$ ($10^{-4}$ Å) | $\tau_1$ (ps) | $A_2$ ($10^{-4}$ Å) | $\tau_2$ (ps) |
|---|---|---|---|---|
| $\Delta(u_{Fe})^2$ | 10.8 ± 2.5 | 0.38 ± 0.08 | 8.9 ± 2.0 | 1.8 ± 0.7 |
| $\Delta(u_{Pt})^2$ | 4.2 ± 3.3 | 0.5 ± 0.2 | 7.3 ± 3.0 | 1.3 ± 0.4 |

**Table 1. Summary of the fit parameters** for the data shown in Fig. 2C using a fit function of the type $A_1(1 - e^{t/\tau_1}) + A_2(1 - e^{t/\tau_2})$.

## B. Wave vector resolved phonon dynamics

Insight into the mechanism by which energy is transferred to the lattice to and from electronic and spin degrees of freedom needs wavevector and time- resolved information on the evolution of the phonon population after laser pulse excitation. The primary source of lattice heating is the hot electron bath that gets excited directly by the laser. The subsequent energy transfer to phonons takes place via electron-phonon scattering events that show a pronounced wavevector dependence, i.e. Brillouin zone boundary phonons often are preferentially populated under the non-equilibrium conditions following ultrafast laser heating [25]. Diffuse electron diffraction has been used as a unique tool to directly probe the wavevector dependence of such transient phonon populations [26]. Here we extend diffuse scattering to the case of FePt with the aim of separating the nonequilibrium motion of Fe and Pt atoms for selected phonon modes.

Figure 3 shows the time-resolved diffuse electron scattering for FePt along the ΓX direction in reciprocal space which corresponds to the a and b crystollographic axes in the inset of Fig. 1C. For FePt we can write the diffuse scattering intensity following ref. [26, 27] as the sum over the phonon modes for each wavevector **k** in the FePt Brillouin zone, i.e.

$$I_{DS}(\mathbf{q}) = \sum_j \frac{1}{\omega_{\mathbf{k},j}} \left(n_{\mathbf{k},j} + \tfrac{1}{2}\right) \left[ \frac{F_{Pt}(\mathbf{q})}{\sqrt{m_{Pt}}} \left(\mathbf{q} \cdot \mathbf{e}_{\mathbf{k},j}^{Pt}\right)^2 \pm \frac{F_{Fe}(\mathbf{q})}{\sqrt{m_{Fe}}} \left(\mathbf{q} \cdot \mathbf{e}_{\mathbf{k},j}^{Fe}\right)^2 \right]^2 \qquad (3)$$

where $m_{Pt,Fe}$ are the atomic masses, $F_{Pt,Fe}(\mathbf{q})$ the electron scattering form factors and $\mathbf{e}_{\mathbf{k},j}^{Pt,Fe}$ the phonon eigenvectors for Pt and Fe atoms, respectively. $\omega_{\mathbf{k},j}$ and $n_{\mathbf{k},j}$ describe energy and occupation number, respectively, for phonons at wavevector $\mathbf{k}$ and branch $j$. The wavevector $\mathbf{k}$ is defined within the Brillouin zone centered on the Bragg peak at a reciprocal lattice vector $\mathbf{q}_{hkl}$. The wavevector transferred in the diffuse scattering process is given by $\mathbf{q} = \mathbf{q}_{hkl} + \mathbf{k}$. Plus and minus signs correspond to even and odd Bragg orders.

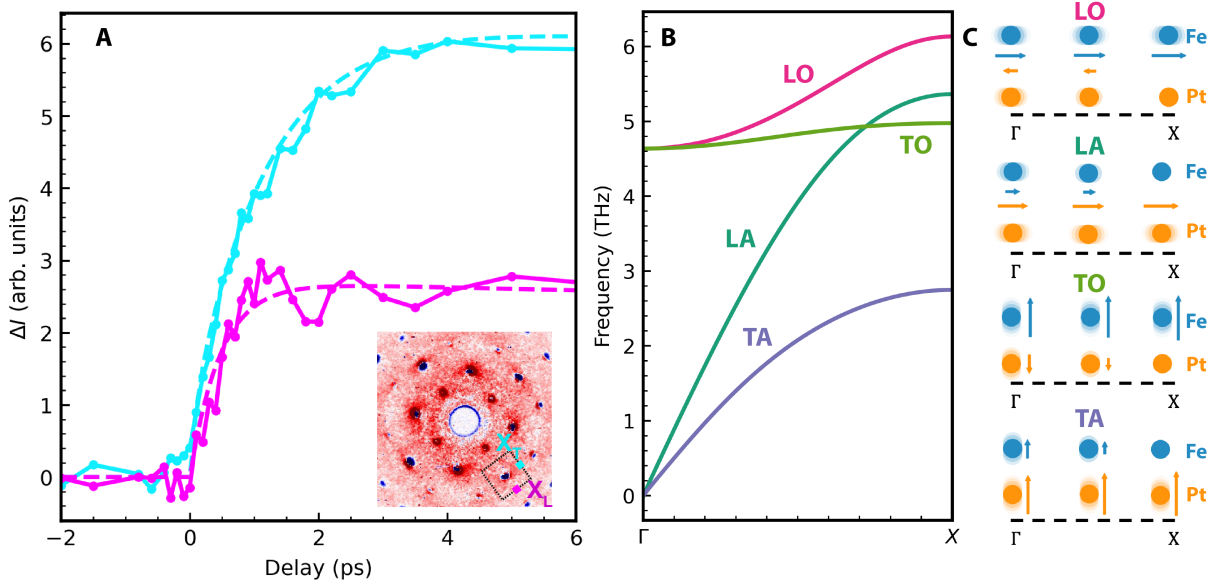

**Fig. 3. Temporal evolution of diffuse scattering**. (**A**) diffuse scattering for two X points around the 200 Bragg peak, the regions of interest have been marked in the inset. (**B**) Phonon dispersions along the ΓX direction for the phonon modes detected in the UED experiment of Fig. 1. C) Illustration of the calculated phonon eigenvectors of the respective modes at three wavevectors along the ΓX direction. Symmetry dictates that at the Brillouin zone boundary X-point only either Fe or Pt atoms are displaced for each mode.

It is straightforward to see from Eq. (3) that reaching the Brillouin zone boundary, for instance, at the X-point from adjacent odd and even Bragg orders requires the zone-boundary phonon modes to have eigenvectors where either Fe or Pt atoms are at rest, i.e. the corresponding $\mathbf{e}_{\mathbf{k},j}^{Pt,Fe}$ must be zero. This is reproduced in the calculations shown in Figs 3B, C that were performed following ref. [25].

The data in Fig. 3A can be described by exponential increases of the form $A(1 - e^{t/\tau})$. We obtain for the $X$-points marked in the inset of Fig. 3A the fit parameters summarized in Table 2 where the term $(\boldsymbol{\varepsilon} \cdot \mathbf{e}_{L,T}^{Pt,Fe})^2$ describes the square of directional cosines for phonons with longitudinal (L) and transverse (T) polarization at the two X-points. This demonstrates that measurements at the $X_L$ and $X_T$ points are sensitive mainly to longitudinal and transverse phonon polarizations, respectively.

|       | $A$ ($10^{-3}$ arb. units)) | $\tau$(ps)    | $(\boldsymbol{\varepsilon} \cdot \mathbf{e}_L^{Pt,Fe})^2$ | $(\boldsymbol{\varepsilon} \cdot \mathbf{e}_T^{Pt,Fe})^2$ |
|-------|------------------------------|---------------|------------------|------------------|
| $X_L$ | 2.7 ± 0.1                    | 0.47 ± 0.06   | 1.0              | 0.0              |
| $X_T$ | 6.0 ± 0.1                    | 0.97 ± 0.4    | 0.07             | 0.9              |

**Table 2. Summary of the fit parameters** for the data shown in Fig. 3A using a fit function of the type $A(1 - e^{t/\tau})$. The terms $(\boldsymbol{\varepsilon} \cdot \mathbf{e}_{L,T}^{Pt,Fe})^2$ describe the directional cosines for the corresponding phonon polarization at regions $X_{L,T}$ marked in the inset of Fig. 3A.

## C. Element-specific lattice expansion

Here we describe an extension of the average lattice expansion for FePt nanoparticles beyond what is shown in Fig. 1 and has been reported so far [10]. We confine ourselves to normal incidence measuring the a, b lattice expansion depicted in Fig. 1B. However, the interference of the scattering amplitudes from Fe and Pt atoms described in Eq. (1) will also allow us to corroborate if the observed lattice expansion is the same for both sublattices or not.

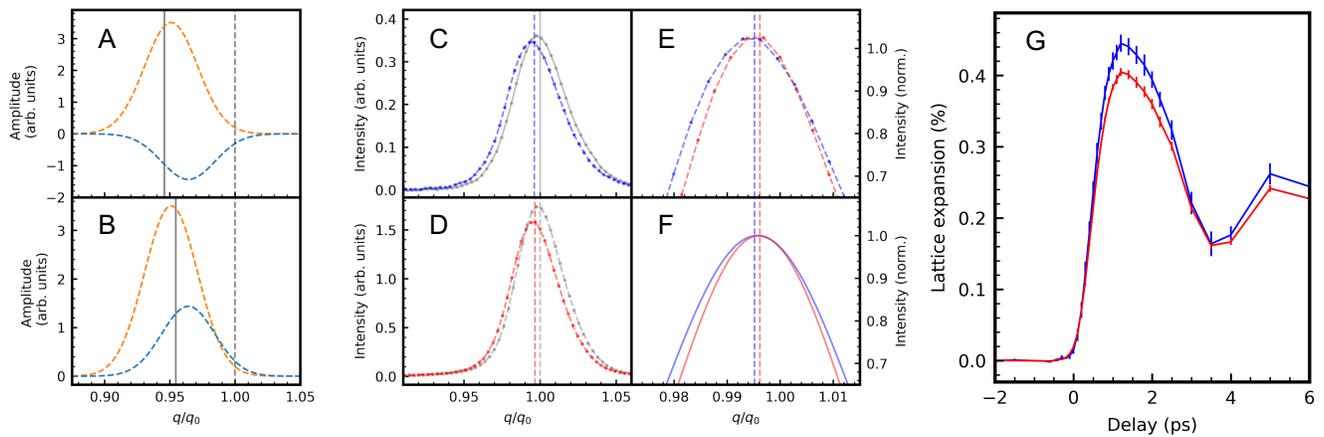

**Fig. 4. Different lattice expansion of Fe and Pt sublattices.** (**A**-**B**) Cartoon of the electric field amplitude scattered at odd (**A**) and even (**B**) Bragg orders when the Fe sub-lattice (light blue) is displaced less than that of Pt (orange). Only when the Fe and Pt lattice spacings are different will the resulting odd and even Bragg peak intensities (marked by the black solid lines) occur at different wavevectors. (**C**) Measured Bragg peak intensity profile for the 300 order before laser excitation (grey) and at a pump-probe time delay of 1ps (blue). (**D**) Measured Bragg peak intensity profile for the 310 order before laser excitation (grey) and at a pump-probe time delay of 1ps (red). (**E**) Comparison of the measured 300 (blue) and 310 (red) Bragg peak profiles close to the peak maxima. (**F**) Comparison of the calculated 300 (blue) and 310 (red) Bragg peak profiles for a Pt sublattice expansion of 0.42% and an Fe expansion of 0.37%. (**G**) Measured lattice expansion for even (red line and symbols) and odd (blue line and symbols) Bragg orders. An extended dataset is shown in Fig. 5.

The situation is schematically depicted in Figs. 4A, B where the scattering amplitudes are illustrated for odd and even Bragg orders, respectively. If the lattice expansion is the same for Fe and Pt sublattices the corresponding odd and even Bragg peaks will be shifted the same amount from the equilibrium lattice position marked by the gray dashed line. If, however, the Pt sub-lattice (blue dashed lines) expands less than the Fe sub-lattice (red dashed lines) the intensity maxima of the interfering scattering amplitudes for odd and even Bragg orders (indicated by black vertical lines in Figs. 4A, B) will no longer match each other. For destructive interference at odd Bragg orders (Fig. 4A) the Bragg intensity maxima will shift to lower values of $q/q_0$ than for the constructive interference at even Bragg orders (Fig. 4B).

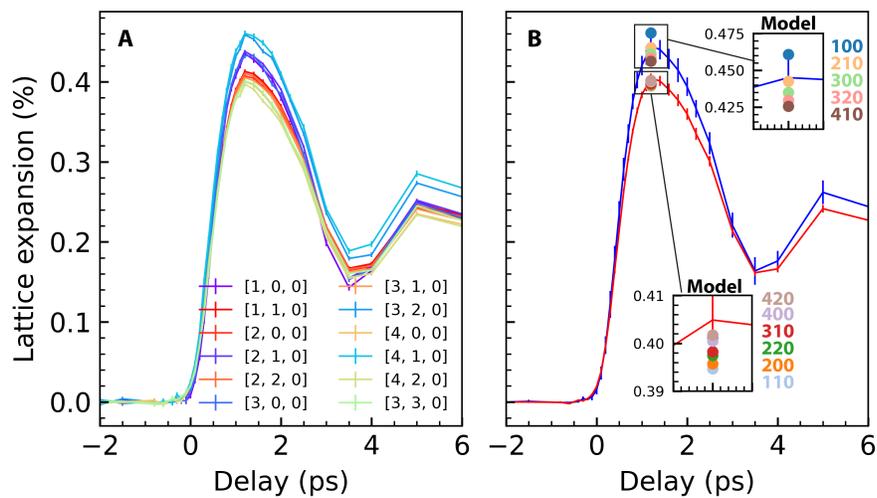

**Fig. 5. Peak positions as a function of time delay.** A) Time-resolved intensities for all measured Bragg peaks. B) Measured lattice expansion for even (red line and symbols) and odd (blue line and symbols) Bragg orders as in Fig. 4G. Calculated intensity maxima of the individual Bragg orders for the model described in the text with a Pt sub-lattice expansion of 0.42% and an Fe sub-lattice expansion of 0.37% are shown as color circles.

Although the shift between odd and even orders is relatively small, it can be clearly seen near the Bragg peak intensity maxima in Figs. 4E and 4F. The peak intensity is normalized to remove Debye-Waller attenuation effects seen in Figs. 4C and 4D and described in section A. We can model the Bragg peak shift observed in Fig. 4E with a Pt sublattice expansion of 0.42% and an Fe expansion of 0.37% as shown in Fig. 4F. An extended dataset together with the predicted intensity maxima is shown in Fig. 5. This leads to an average peak shift of all measured odd (red line and symbols) and even (blue line and symbols) Bragg peaks displayed in Figs. 4G and 5B.

IV. DISCUSSION

There are several aspects related to FePt nanoparticles that make them unique candidates to study the non-equilibrium interplay between electronic, magnetic and lattice degrees of freedom. While optical and X-ray pump-probe studies have focused on electron and spin thermalization times following laser heating [10, 28] our observation of a unit-cell volume expansion in Fig. 1C highlight the intricate link behind these processes. The observed timescale for this lattice expansion indicate it is driven by acoustic strain waves. We have previously observed the longitudinal acoustic (LA) phonons that form the coherent phonon wave packets [29] driving this expansion in analogy to the THz strain wave propagation in ultrathin Fe films [30]. Using the known speed of sound for LA phonons of 4.6 nm/ps [29] together with the 7 nm nanoparticle diameter, we find that the lattice expansion risetime of ~0.8 ps in Fig. 1C corresponds to the strain wave traversing about half of a nanoparticle. This is a reasonable estimate since such strain waves originate at the nanoparticle perimeter and propagate inwards leaving an expanded lattice behind [30].

The unit cell volume increase by more than 0.4% (see Fig. 1C) could also significantly alter the electronic structure and affect electron-phonon coupling and energy transfer as observed previously for Ni [31]. We should therefore look for evidence in this direction for the FePt nanoparticle system. An obvious candidate is the fast timescale observed in the longitudinal phonon populations probed in Fig. 3A at the $X_L$-point. We determined the rise time constant to 0.47 ± 0.06 (see Table 2), however, a saturation-like leveling off is observed at longer times closer to that also evident in the unit cell expansion data (Fig. 1C). It is, therefore, conceivable that the two effects are linked. Inspection of the phonon dispersions close to the $X$-point in. Fig. 3B shows that two longitudinal modes can be detected in our UED geometry, one optical (LO) and one acoustic (LA) mode. They both reach relatively similar frequencies at the $X$-point, however, for the LO mode only Fe atoms vibrate while the LA mode is characterized by only Pt vibrations (see Fig. 3C). From our measurements alone we cannot differentiate if one of the two modes is preferentially occupied, however, calculations in ref. [25] favor a stronger electron-phonon coupling and, thus, mode occupation for the LO mode. Such an assignment with the correspondingly stronger Fe vibration amplitude would also agree with the observed initial increase of the mean square displacements especially for Fe atoms in Fig. 2C that occurs on a similar timescale.

On longer timescales beyond 1 ps we observe a slower increase in the mean square displacements in Fig. 2C as well as the population of a X-point phonon mode with transverse polarization in Fig. 4A. Both could be related and are caused by a reduced electron-phonon coupling, possibly in part due the the unit cell expansion. While we cannot rule out from Fig. 3 that the LO mode becomes populated and contributes to the diffuse scattering signal, the larger density of states observed for the TO mode seems to give this mode the preference to contribute at least to the mean square displacement increase in this time range.

The observed unequal expansion of Fe and Pt sub-lattices may be traced back to the strain-induced lattice expansion of FePt nanoparticles. Strain waves propagate with the speed of sound for LA phonons. In FePt this is about 4.6 nm/ps [29] which reproduces the observed oscillation period of the volume-conserving breathing mode in Fig. 1D for 7-8 nm diameter particles. However, the first such oscillation cycle will be influenced and is in fact driven by electronic and magnetic stresses [10, 11, 30]. Such stresses are related to the non-equilibrium heating of electrons and reduction of the magnetic order that occurs on timescales of just a few 100 fs, i.e. within the transit time of LA strain waves through the nanoparticles. While a detailed modeling of these processes is beyond the scope of the present paper, it is straightforward to imagine that this can lead to an inhomogeneous lattice expansion across a nanoparticle around the 1 ps pump-probe delay time where the maximum a,b-axis lattice expansion is observed (see Figs. 1 D and 4G). The stronger average lattice expansion for the Pt sublattice (of 0.42%) compared to that of Fe (0.37%) is in line with a larger wheight of Pt than Fe to the eigenvectors of this LA mode throughout the Brillouin zone (Fig. 3C).

V. SUMMARY AND CONCLUSIONS

In this work we have developed a novel approach to study the element specific lattice dynamics of FePt nanoparticles. It is based on using the constructive and destructive interference effects present in multi-atomic lattices. We show that these effects can be utilized based on the simultaneous access to multiple Bragg peaks in ultrafast relativistic electron diffraction. We demonstrated that the method allows us to separate Fe and Pt mean square displacements and corroborated the assignment with diffuse electron diffraction measurements. We identified an inhomogeneous ultrafast lattice expansion that is larger for the Pt than the Fe sublattice, possibly driven by coherent longitudinal acoustic phonon wave packets.

ACKNOWLEDGMENTS

The UED work was performed at the SLAC MeV-UED, which is supported in part by the DOE BES SUF Division Accelerator & Detector R&D program, the LCLS Facility, and SLAC under contract Nos. DE-AC02-05-CH11231 and DE-AC02-76SF00515. DT and HAD acknowledge support by the Swedish Research Council (VR). HAD acknowledges support by the Knuth and Alice Wallenberg Foundation (KAW). XJW and KST acknowledge support by the Deutsche Forschungsgemeinschaft (DFG, German Research Foundation) through the Collaborative Research Centre (CRC) 1242 (project number 278162697, project C01 Structural Dynamics in Impulsively Excited Nanostructures)".

AUTHOR DECLARATIONS

Conflict of Interest: The authors have no conflicts to disclose.

DATA AVAILABILITY

The data that support the findings of this study are available from the corresponding author.